\documentclass[reprint,superscriptaddress]{revtex4-1}
\usepackage{amsmath,amssymb,units,amsfonts}
\usepackage[dvipdfmx]{graphicx}
\usepackage{color}
\usepackage{verbatim}
\usepackage{tikz}
\usepackage{siunitx}
\usepackage{gensymb}
\usepackage{bm}

\begin{document}
\title{Bottom-up approach to room temperature quantum systems}

\author{Bochao Wei}
\affiliation{School of Physics, Georgia Institute of Technology, 837 State St, Atlanta, Georgia 30332, USA}
\author{Chao Li}
\thanks{Present address: Research Laboratory of Electronics, Massachusetts Institute of Technology, Cambridge, Massachusetts 02139, USA.}
\author{Ce Pei }
\author{Chandra Raman}
\email {Corresponding author:  craman@gatech.edu}
\affiliation{School of Physics, Georgia Institute of Technology, 837 State St, Atlanta, Georgia 30332, USA}

\begin{abstract}
\noindent {\bf Abstract.} We demonstrate a key ingredient in a “bottom-up” approach to building complex quantum matter using thermal atomic vapors.  We have isolated and tracked very slowly moving individual atoms without the aid of laser cooling.  Passive filtering enabled us to carefully select atoms whose three-dimensional velocity vector has a magnitude below $\bar{v}/20$, where $\bar{v}$ is the mean velocity of the ensemble.  Using a novel photon correlation technique, we could follow the three-dimensional trajectory of single, slowly moving atoms for   $> 1\mu$s within a $25\mu$m field of view, with no obvious limit to the tracking ability while simultaneously observing Rabi oscillations of these single emitters.  Our results demonstrate the power and scalability of thermal ensembles for utilization in quantum memories, imaging, and other quantum information applications through bottom-up approaches.

\end{abstract}
\maketitle

\begin{figure}[ht]
\centering
\includegraphics[width=\linewidth]{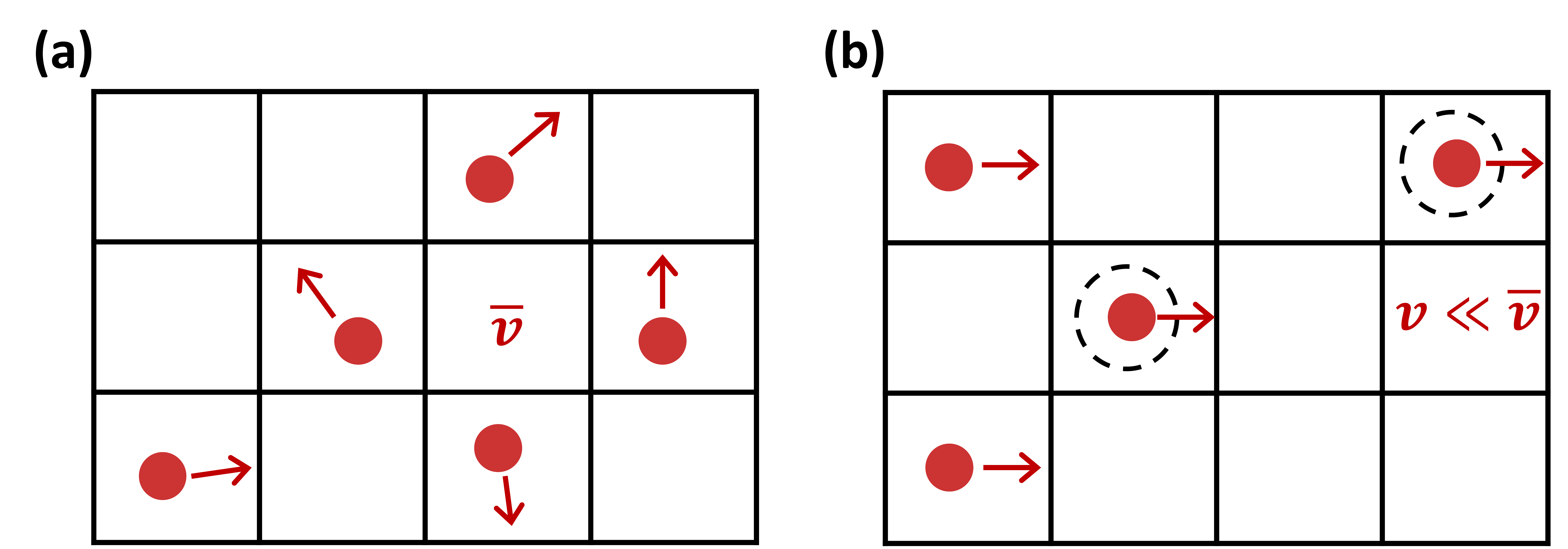}
\caption{Concept of the bottom-up approach to room temperature quantum information processing with neutral atoms under a high-resolution microscope.  The field of view is divided into mesoscopic cells, with less than one atom per cell on average.  (a) Ordinary vapor with randomly oriented velocities of magnitude $\bar{v}$. (b) Three-dimensional velocity selection with $v\ll \bar{v}$.  Atoms can be tracked from one cell to the next.  This constitutes a new paradigm for a bottom-up approach to quantum information processing.     
}
\label{fig:0} 
\end{figure}

In recent years there has been a surge of interest in room-temperature atomic vapors for applications in quantum information science.  In contrast to laser-cooled samples, they are straightforward to fabricate, highly scalable, and can be operated continuously.    One can define two broad thrusts to this research--a ``top-down'' and a ``bottom-up'' approach.  The former, which has been adopted in four-wave mixing experiments \cite{shu2016subnatural,zhao2019direct,becerra2008nondegenerate,main2021preparing}, seeks to engineer collective quantum behavior within the vapor and ignores the discrete nature of the constituent particles. Analogous behavior with ultracold atoms are collective excitations such as phonons and magnons \cite{fukuhara2013microscopic,ozeri2005colloquium,stamper2013spinor}.  The latter approach seeks to construct a complex quantum system from individual atomic building blocks.  This approach has been followed for laser-cooled atoms in optical tweezer arrays \cite{wilson2022trapping,kaufman2021quantum}, and trapped ions \cite{bruzewicz2019trapped,singer2010colloquium}, but is  completely undeveloped for thermal atomic vapors.  At issue is the rapid and random thermal motion of the atoms that makes it difficult to track them (see Fig.\ \ref{fig:0}a).  If one could address this issue, the possibilities are clearly enormous--a typical rubidium vapor cell at 100 $\degree C$ contains $O(10^9)$ completely indistinguishable quantum systems within a 1 mm$^3$  volume.  Even a small fraction of such a large ensemble constitutes a huge and readily available resource for quantum information if it can be harnessed. 

Fig. \ref{fig:0} illustrates an array of ``mesoscopic'' cells within a thermal vapor.  The array need only be partially ordered to be useful, provided one knows the  population of each cell accurately.  This can be determined with a high-resolution microscope and single photon detection.  Individual cells are taken to be  $25\ \mu$m in size for convenience. 

Around 1600 cells can be constructed within a field of view $\sim 1 \times 1$ mm.  However, at typical thermal velocities of 300 m/s, atoms cannot be observed for more than $\sim 83$ ns within one cell, which is too short for most purposes. Moreover, atoms move in random directions and cannot be tracked.  

\begin{figure}[t]
\centering
\includegraphics[width=\linewidth]{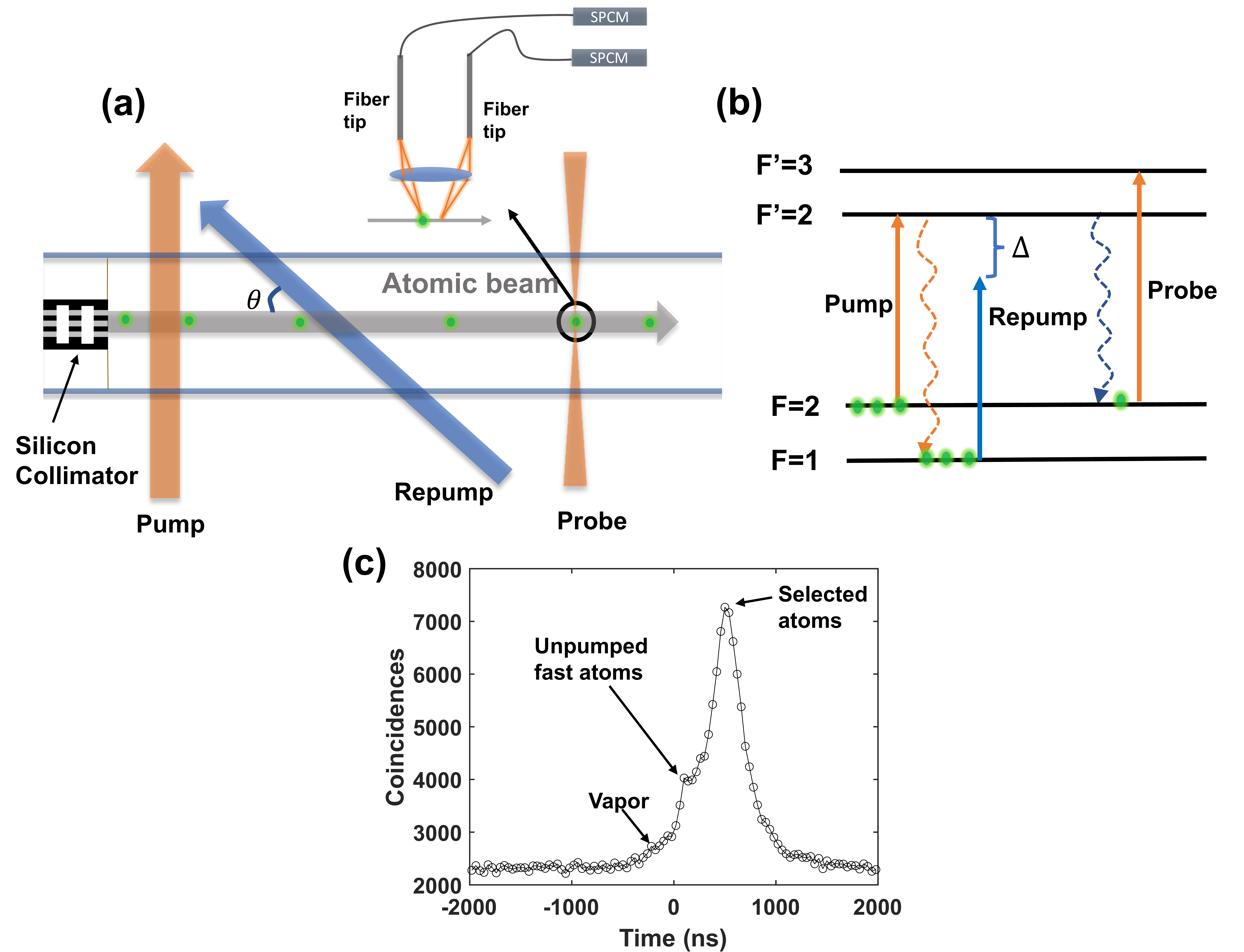}
\caption{Selecting and measuring single slow atoms. (a) Schematic of the experiment. Atoms from the silicon collimator are first pumped into $F=1$ and then selectively pumped back to $F=2$. A Doppler-free probe beam is used to detect the selected atoms. The angle between repump and the atomic beam is $\theta=$47 \degree. Two cleaved fiber tips are placed on the image plane of the imaging system with a numerical aperture (NA) of 0.42. (b) The involved energy levels. The repump is detuned by $\Delta$ to select a certain group of atoms. (c) The raw coincidences with different time delays from two SPCMs at $\Delta=-80$ MHz, corresponding to a velocity of $v\approx 100$m/s.}
\label{fig:1} 
\end{figure}

In this work, we address this issue and experimentally demonstrate the feasibility of such a “bottom-up” approach to quantum information science with thermal vapors. The key advance we have made is to isolate a sub-ensemble of atoms whose three-dimensional velocity vector is 20 times smaller in magnitude than the mean, which extends the observation time to $> 1$ $\mu$s and simultaneously enables tracking of atoms across cells, since all atoms travel in the same direction. As a first step to constructing bottom-up thermal quantum systems, we observed quantum-mechanical antibunching and the correlated photon emission from a single atom for $>1\ \mu s$. We also observed large values of the second order  $g^{(2)}(\tau)$ and third-order coherences  $g^{(3)}(\tau_1,\tau_2)$. This indicates its potential to be a simple source of photon pairs or triplets for quantum applications. 


The schematic of the experimental setup is shown in Fig \ref{fig:1} (a).
We employ a miniature atomic beam generation device based on the chip-scale cascaded collimator \cite{li2019cascaded}, a two-dimensional passive collimation device. In this design, off-axis atomic beam components are kept within the source region, greatly increasing the collimation and suppressing the vapor component, which is key to this experiment.  This miniature device could be inserted directly into a 12 mm $\times$ 12 mm $\times$ 42 mm cuboid shape glass vacuum cell.
The collimator consists of 20 channels, each with a cross-section of 100 $\mu$m $\times$100 $\mu$m, resulting in a beam with a narrow divergence angle ($\theta_{1/2}=0.013$ rad, corresponding to a transverse velocity spread of only $\pm 4$ m/s). 

By using an atomic beam, we could select atoms whose longitudinal velocity is substantially lower than the average.  The selection was performed upstream of the detection.  When combined with the two-dimensional passive filtering within the source, this isolates atoms with a small 3-dimensional velocity vector.  To achieve single emitter dynamics, we ensure that the mean number of atoms in the detection region $\langle N \rangle \ll 1$ by reducing the oven temperature.  




The $^{87}$Rb $D_2$ line transition diagram and experimental procedure are shown in Fig \ref{fig:1} (b).
To select the slow atoms, a Doppler-free pump beam first pumps all atoms into the hyperfine $F=1$ state, while an angled repump beam then selectively pumps atoms back to $F=2$ depending on their longitudinal velocity. Finally, a Doppler-free probe beam perpendicular to the atomic beam couples the $F=2 \rightarrow F'=3$ transition to detect the selected atoms. The selected atoms are expected to have a velocity center at $v_c=\frac{-\Delta}{k\cos{\theta}}$ where $\Delta<0$ and $k$ are the detuning and the wave number of the repump beam, and $\theta$ is the angle between repump and the atomic beam.  Low velocities were selected by decreasing the detuning $|\Delta|$ toward $0$.

The ratio of atoms with a velocity below 30 m/s in a thermal rubidium atomic beam is very small ($\approx 10^{-5}$). Normal Doppler-sensitive spectroscopy cannot observe these atoms, as it will be limited by electronic noise, vapor contributions, and the fluorescence from $^{85}$Rb. Here we demonstrated a  single-atom photon correlation method to track their motion across two cells defined in Fig. \ref{fig:0}b. This technique works as follows. Two bare fibers are cleaved and fixed on a plastic holder to keep their distance at 450 $\mu$m. This holder is then fixed on the image plane of the microscope and forms two detection regions separated by $d\approx$ 55 $\mu$m in the plane of the atoms, as shown in Fig \ref{fig:1}(a).  The other end of each fiber is connected to a single photon counting module (SPCM) where the detected photons are time tagged and analyzed. 

When a single atom passes through two detection regions, the photons collected from the two fibers will contribute to time-ordered coincidences with a delay $\tau=\frac{d}{v}$. The accidental coincidences from laser scattering, detector dark counts, etc., do not depend on time delay and can be subtracted later. The focused probe beam has a beam diameter of $2w\approx$ 120 $\mu$m that overlaps both detection regions. Figure \ref{fig:1} (c) shows the raw coincidence data for $\Delta=-80$ MHz. The peak around +500 ns is the contribution from velocity-selected atoms with a center velocity $v=\frac{d}{\tau}\approx100$ m/s. Some fast atoms in the atomic beam escape the pumping process and contribute to the small bump at around +100 ns. During the data acquisition time, a small rubidium vapor gradually builds up and contributes to coincidences with both positive and negative time delays around zero.  This vapor could be removed in future experiments for greater selectivity, for example, by adding a small amount of graphite to the vacuum cell.

We can analyze the second-order temporal coherence between SPCM A and SPCM B: 
\begin{equation}
   g_{AB}^{(2)}(\tau)=\frac{\langle I_A(t)I_B(t+\tau)\rangle}{\langle I_A(t) \rangle\langle I_B(t+\tau) \rangle} 
\end{equation}
$g_{AB}^{(2)}(\tau)$ measures the distribution of coincidences with time delay $\tau$ and $g_{AB}^{(2)}(\infty)\rightarrow 1$ represents accidental coincidences. The correlated part $g_{AB}^{(2)}(\tau)-g_{AB}^{(2)}(\infty)$, after normalization, is the coincidence probability density in the time domain $n_{AB}(\tau)$. 
Given $n_{AB}(\tau)d\tau=n_{AB}(v)dv$ and $\tau=\frac{d}{v}$, we can derive $n_{AB}(v)$, which is the number density of coincidences contributed by atoms whose velocity is $v$.  
The coincidences generated by each atom are proportional to the square of transit time through a single fiber's detection region, whose diameter is $d_f$.  Then the atom probability density $\rho(v)$ for the flux is derived from coincidence data by using $n_{AB}(v)\propto\rho(v)\cdot\frac{d_f^2}{v^2}$ (check Supplement for details).

\begin{figure}[t]
\centering
\includegraphics[width=\linewidth]{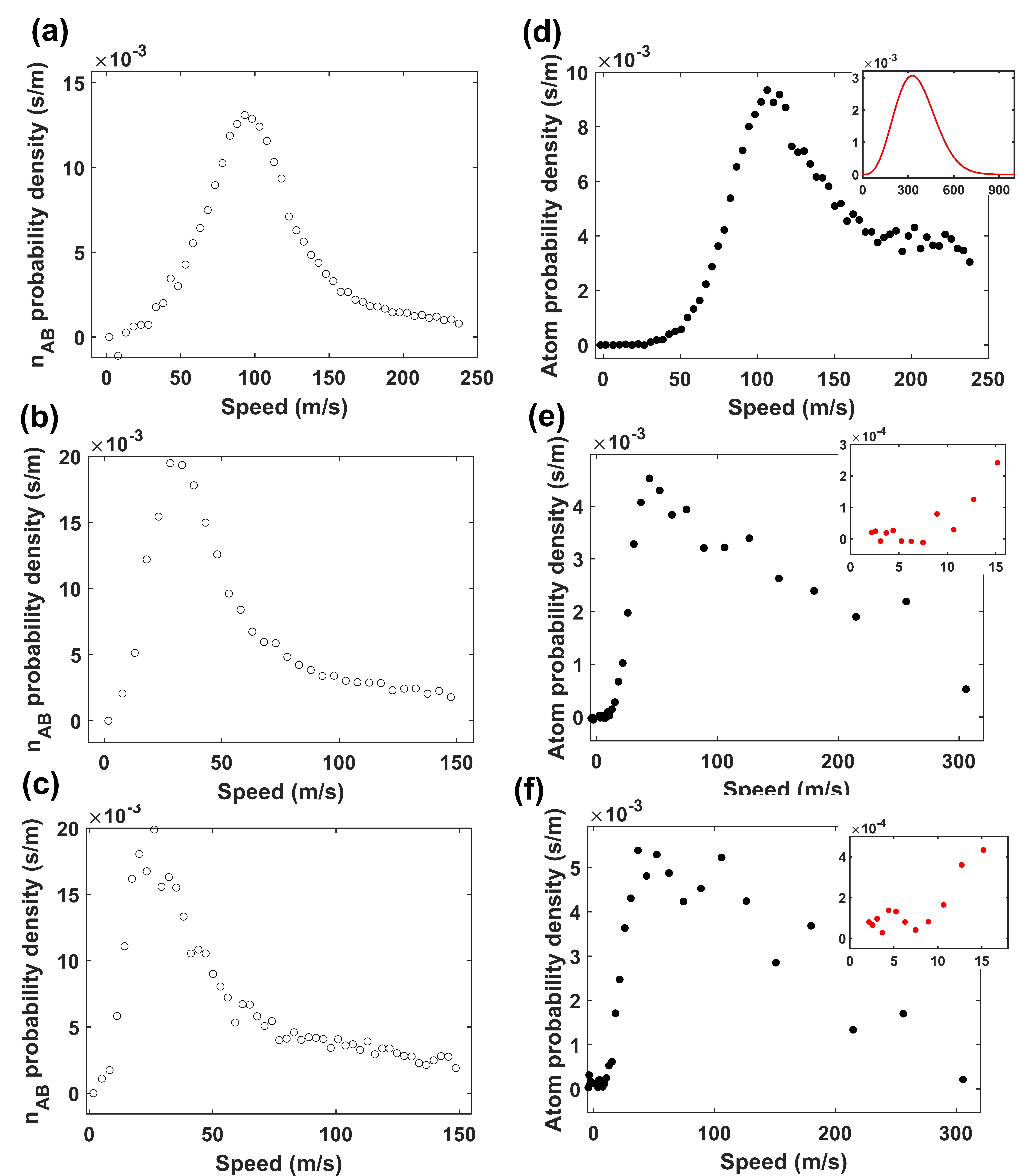}
\caption{Atomic velocities measured by correlations.  (a-c) on the left are the coincidence distributions in the velocity domain for detunings $\Delta=-80$ MHz,$-20$ MHz, and $-10$ MHz, respectively. The peak velocity shifts from 93 m/s to 30m/s to 20 m/s.  (d-f) on the right are the corresponding atom probability density distributions. In (d), the peak velocity is 106 m/s, while the inset shows the original thermal distribution at 100 $\degree$C where the peak is much higher, at 300 m/s.  The insets in (e) and (f) are  zoomed-in plots of the low-velocity region where atoms with velocities around 15m/s can be clearly distinguished.  }
\label{fig:2} 
\end{figure}

For $\Delta=-80$ MHz and oven temperature 100 $\degree$C, the data for $n_{AB}(v)$ is shown in Fig \ref{fig:2}(a). It shows the photon coincidences contributed by atoms with velocities from 0 to 250 m/s. Fig \ref{fig:2}(d) is the calculated atom probability density distribution $\rho(v)$. Compared with the original thermal atomic beam velocity distribution (Fig \ref{fig:2} (d) inset), the selected atoms have a much lower velocity--the peak is at 106 m/s, which agrees reasonably well with the theoretical expectation of 92 m/s.

To select even slower atoms, we used $\Delta=-20$ and $-10$ MHz, whose data for $n_{AB}(v)$ are shown in Fig \ref{fig:2}(b) and (c). We can clearly see the coincidences have shifted to lower velocities, with the peak occurring at 30 m/s and 20 m/s, respectively. The corresponding $\rho(v)$ are shown in Fig \ref{fig:2} (e) and (f). Compared with the coincidence $n_{AB}(v)$, the atom probability density distribution is broader and has bigger tails in high velocities. The reason is that slower atoms contribute more coincidences. Thus, the peak locations for $n_{AB}(v)$ are closer to 0, and the peaks are narrower.  The expected peak of atom probability density is at 23 m/s and 12 m/s for  Fig \ref{fig:2} (e) and (f), while the actual peak locations are both at around 50 m/s. 

Several nonidealities limited the velocity selection purity. 
The imperfection in imaging can cause a small probability of detecting photons from atoms between two fiber tips, creating spurious coincidences similar to ultra-fast atoms. 
Some fast atoms managed to avoid being optically pumped through the pump beam, and the background rubidium vapor within the small glass chamber increased with time during the experiment. 
After averaging for several hours, the correlation method we used could distinguish the small correlated signals, but some faster atoms inevitably shifted the peak location and cause the long tail in Fig \ref{fig:2} (e) and (f). In spite of these nonidealities, atoms with velocities around 15 m/s can clearly be observed, as shown in the inset of Fig \ref{fig:2} (e) and (f). This demonstrates that we can isolate and directly observe slow atoms with a velocity $< \bar{v}/20$, where $\bar{v}$ is the mean velocity of the unselected atomic beam. 

 In the remainder of this paper, we show the experimental measurements on single atoms in an atomic beam, demonstrating their utility in a ``bottom-up'' approach to quantum systems. 
 \begin{figure}[t]
\centering
\includegraphics[width=\linewidth]{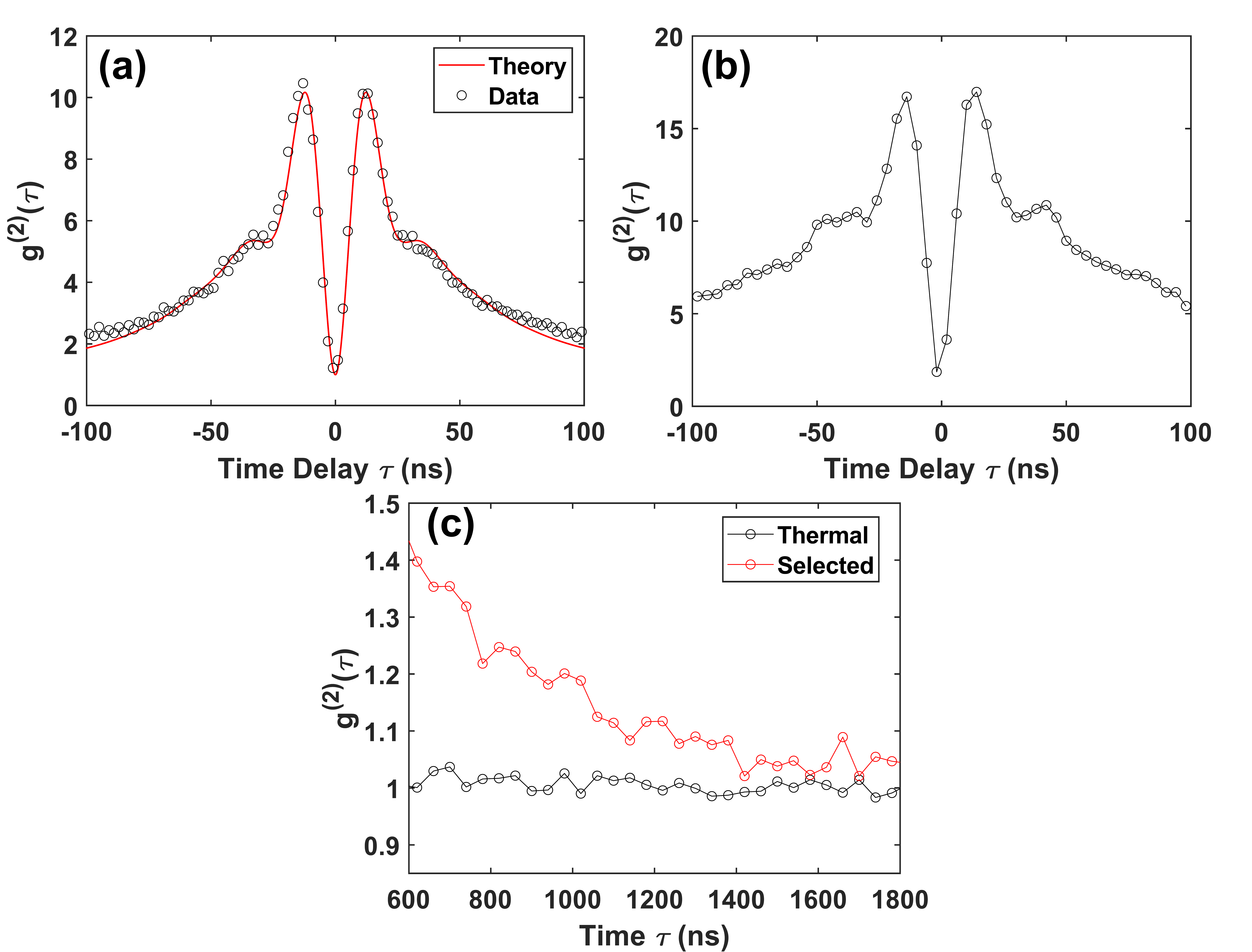}
\caption{(a) $g^{(2)}(\tau)$ with thermal atomic beam at 78 $\degree$C. The time bin size is 2 ns. Red line is the theory curve with $\langle N\rangle=0.138$, $L=25$ $\mu$m. Check the supplement for derivation. (b) $g^{(2)}(\tau)$ with selected atoms at 100 $\degree$C. The selected atoms are more confined in space and suffer less intensity variance in the probe beam. Thus, the second Rabi peak is more visible. The time bin size is 4 ns. (c) $g^{(2)}(\tau)$ for thermal and selected atoms are plotted together with long time delays.}
\label{fig:3} 
\end{figure}
A key signature of single atoms is the photon antibunching effect \cite{kimble1977photon}. In order to measure the second-order correlation function $g^{(2)}(\tau)$,
the collector with two fiber tips is replaced by a single fiber tip that is connected to a 50:50 fiber splitter and two SPCMs to achieve a Hanbury-Brown and Twiss configuration. The field of view has a  diameter $d_f\approx 25\mu$m.

We first measured the $g^{(2)}(\tau)$ of an unfiltered thermal atomic beam by removing the pump and repump beams and reducing the oven temperature to 78 $\degree$C to achieve $\langle N \rangle <1$ in the field of view. The data are shown in Fig \ref{fig:3} (a).  For classical light the condition $g_2(\tau)\leq g_2(0)$ must be met \cite{loudon2000quantum}, and therefore the observed dip around $\tau=0$ is the evidence for the quantum-mechanical antibunching effect from single atoms \cite{kimble1977photon,kimble1978multiatom}.  After an emission event, an atom needs time to be re-excited to emit a second photon, and therefore the maximum of $g^{(2)}(\tau)$ occurs around the first half Rabi cycle.  At zero time delay, $g^{(2)}(0)=1$ rather than $0$ because the atomic beam follows the Poisson distribution, and the single emitter condition is not always satisfied. We will return to this point shortly.

In comparison with trapped atom systems \cite{diedrich1987nonclassical,gerber2009intensity,tey2008strong}, the peak value of $g^{(2)}(\tau)$ observed at $\tau = \tau_{\max}\approx$ 12 ns was much larger, as high as 10 for the unfiltered thermal atom data.  Such a large value is comparable to what has been observed for correlated photon pairs using four-wave mixing in vapor cells \cite{shu2016subnatural,du2007four}.  This is because the accidental coincidences scale with $\langle N\rangle^2$, while the correlated coincidences scale with $\langle N\rangle$.  Thus in an atomic beam where $\langle N\rangle\ll1$, the ratio of correlated coincidences is much higher. With the Poisson process averaging and transit time correction,
 the $g^{(2)}(\tau)$ for a thermal atomic beam can be written as (see Supplement for derivation):
\begin{equation}
g^{(2)}(\tau)=\xi(\tau) \cdot \frac{g^2_{single}(\tau)}{\langle N\rangle}+1
    \label{eq:g2}
 \end{equation}
where $L$ is the length of the field of view in the atomic beam direction, $\rho(v)$ is the atomic beam Maxwell-Boltzmann velocity distribution, 
and $g^{(2)}_{single}(\tau)$ is the second-order coherence function of a single stationary atom. The full expression for the transit time correction factor $\xi(\tau)$ is given in our supplement.
We can learn two things from Eqn.\ (\ref{eq:g2}).  One is that $g^{(2)}(\tau)-1$ is inversely proportional to the average atom number $\langle N \rangle$ and the high $g^{(2)}(\tau)$ value only appears when $\langle N \rangle<1$. Secondly, the small ratio $g^{(2)}(0)/g^{(2)}(\tau_{\max})\approx 0.1$ indicates a high purity of single atom emission and low contamination by multi-atom events.  
For an ideal single emitter, this ratio is zero.



To confirm and compare this effect, the velocity selection scheme was used to measure the $g^{(2)}(\tau)$ for slow atoms at a repump detuning $\Delta=-20$ MHz, with the data shown in Fig \ref{fig:3} (b). The $g^{(2)}(\tau)$ peak is even higher, reaching 17, due to the smaller averaged atom number for this data, with $g^{(2)}(0)/g^{(2)}(\tau_{\max})\approx 0.06$. Since the transit time for slow atoms is much longer, $g^{(2)}(\tau)$ also decays more slowly at long $\tau$. Figure \ref{fig:3}(c) shows the comparison between the $g^{(2)}(\tau)$ of thermal atoms and selected atoms. The correlated photons can be seen for $\tau>1000$ ns, coming from atoms with $v<25$ m/s. This shows that by using simple velocity selection with thermal beams, we can observe a single atom for longer than 1 $\mu$s. 

The time delay between the photon pairs can be tuned by the probe laser intensity. With the thermal beam at 78 $\degree$C and a probe laser power of 7 $\mu$W, the collected photon pairs have a rate of 0.16 pairs per second per fiber.  While this is a small rate, it can be improved, and in this regard, the simplicity and small size of the source should not be overlooked.  To achieve a practical output flux, one can readily multiplex the output of several cells, for example, by adding more fibers and by improving the collection efficiency, as the pair rate scales quadratically with this quantity.  Moreover, unlike spontaneous parametric down conversion sources \cite{baek2008spectral,couteau2018spontaneous}, this system requires no wavelength filtering and is ideally suited for interaction with rubidium atoms.


Photon triplet generation has been a longstanding challenge in the quantum optics field \cite{khoshnegar2017solid,krapick2016chip, hubel2010direct}. 
We expect our system to also generate photon triplets. The third-order correlation function    $g^{(3)}(\tau_1,\tau_2)$ measures the temporal correlation of three photons:
\begin{equation}
    g^{(3)}(\tau_1,\tau_2)=\frac{\langle I_A(t)I_B(t+\tau_1)I_C(t+\tau_2)\rangle}{\langle I_A(t) \rangle\langle I_B(t+\tau_1) \rangle \langle I_C(t+\tau_2) \rangle} 
\end{equation}
A high $g^{(3)}(\tau_1,\tau_2)$ value means a high probability of detecting three photons with time delay $\tau_1$ and $\tau_2$ compared with other time delays. $g^{(3)}$ was measured using two detectors in the Hanbury Brown and Twiss configuration.  We recorded the arrival time of photons from SPCM A and SPCM B with an accuracy of 350 ps and a dead time $\Theta\approx$ 45 ns. Then, since photons don't distinguish SPCM A and SPCM C, the time tags from SPCM A were used as the time tags for SPCM C. We removed the spurious coincidences at $\tau_2\approx0$ and a partial function $g^{(3)}(\tau_1,\tau_2>\Theta)$ was measured.  
Due to detector dead time, our measurements were sensitive only to the bunching of the triplets  $g^{(3)}_{\max}>1$ occurring at finite delays and not to the antibunching effect near zero. 
\begin{figure}[t]
\centering
\includegraphics[width=\linewidth]{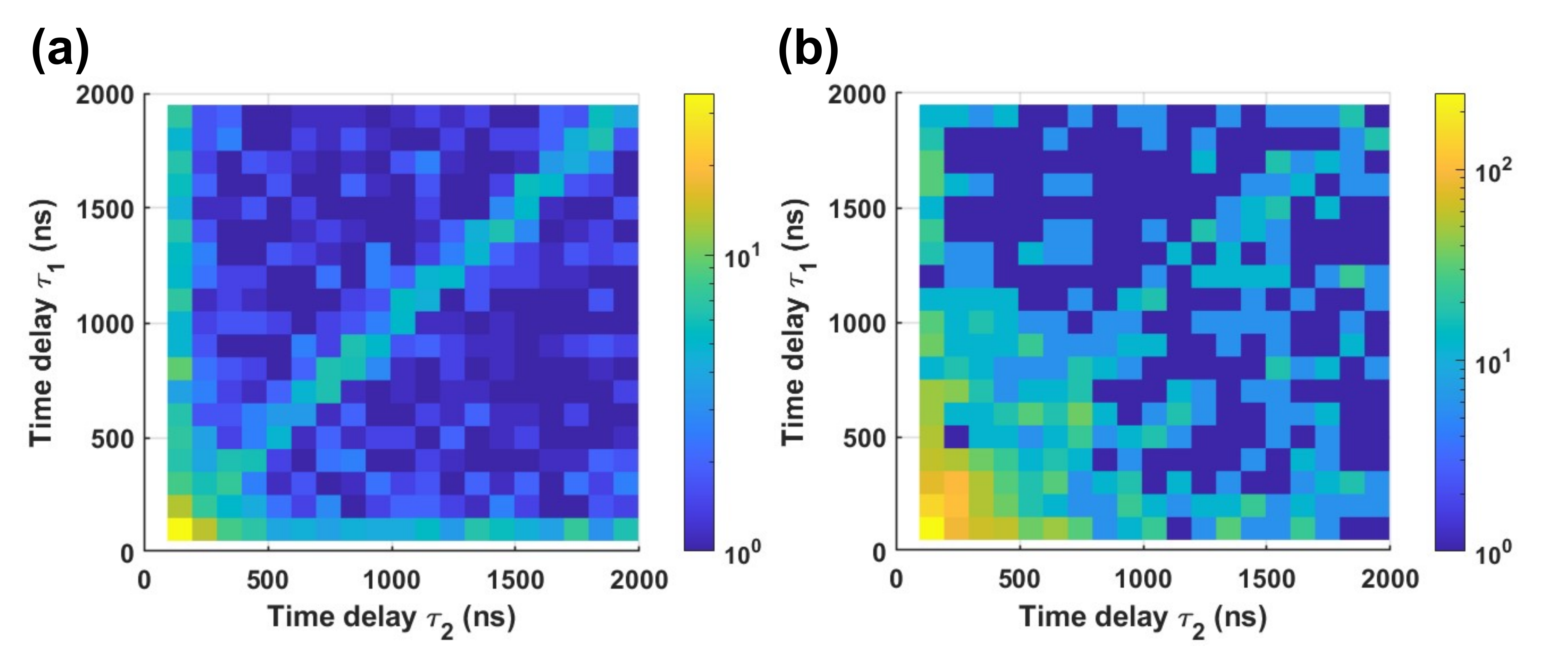}
\caption{Measured $g^{(3)}(\tau_1,\tau_2>\Theta)$. (a) Thermal atomic beam at 78 $\degree$C. The maximum $g^{(3)}$ is around 39. (b) Selected atoms by using a detuning of $\Delta=-20$ MHz. The maximum $g^{(3)}$ is around 280.}
\label{fig:4} 
\end{figure}

Figure \ref{fig:4} (a), (b) shows the data for the thermal atomic beam and $\Delta=-20$ MHz selected atoms, respectively. The time bin size is 100 ns $\times$ 100 ns to reduce the shot noise and $g^{(3)}(\tau_1,\tau_2<\Theta)$ is left blank.
The peak around zero results from the consecutive three photons emitted during the transit of single atoms. 
When $\tau_1\approx\tau_2$, channels B and channel C will have more coincidences as shown in the $g^2(\tau)$ measurements, resulting in a higher value of three-photon coincidences and a diagonal line in Fig. \ref{fig:4} (a). When $\tau_1$ or $\tau_2$ close to zero, the same reason leads to the brighter lines close to the axis. For Fig.\ref{fig:4} (b), the number of three-photon coincidences is not large enough, and this pattern is blurred by shot noise. Comparing Fig. \ref{fig:4}(b) to (a), stronger third-order correlations from slow atoms are detected in large time delays, showing the capability to collect photon triplets from a single atom for more than 1 $\mu$s.

The maximum values of $g^{(3)}(\tau_1,\tau_2>\Theta)$ reach 39 and 280 for each case, showing great potential as a photon triplet source. For the thermal beam, we collected $\approx$0.166 triplets per minute. The rate of photon triplet is proportional to the cubic of collecting efficiency. Improving the collecting efficiency and adding more fibers in the imaging plane can create bright, narrow linewidth photon triplets that are compatible with Rubidium-based systems.

In summary, we have isolated and detected single slow atoms within a thermal atomic beam and measured their unique photon statistics, showing the possibilities inherent in a ``bottom-up'' approach to thermal quantum systems. Improved velocity selectivity can be achieved in the future by using, for example, a two-photon Raman transition for pumping.

\medskip
We acknowledge funding from the Air Force Office of Scientific Research under grant no. FA9550-19-1-0228.
\bibliography{references}
\end{document}


\title{Supplementary material for }

\author{Bochao Wei}
\affiliation{School of Physics, Georgia Institute of Technology, 837 State St, Atlanta, Georgia 30332, USA}
\author{Chao Li}
\thanks{Present address: Department of Electrical Engineering and Computer Science, Massachusetts Institute of Technology, Cambridge, Massachusetts 02139, USA.}
\author{Pei Ce}
\author{C. Raman}
\email {Corresponding author:  craman@gatech.edu}
\affiliation{School of Physics, Georgia Institute of Technology, 837 State St, Atlanta, Georgia 30332, USA}
\maketitle

\renewcommand{\theequation}{S\arabic{equation}}
\renewcommand{\thefigure}{S\arabic{figure}}
\renewcommand{\bibnumfmt}[1]{[S#1]}
\renewcommand{\citenumfont}[1]{S#1}

\section{Experimental set up}
\begin{figure*}[bh]
\centering
\includegraphics[width=0.75\linewidth]{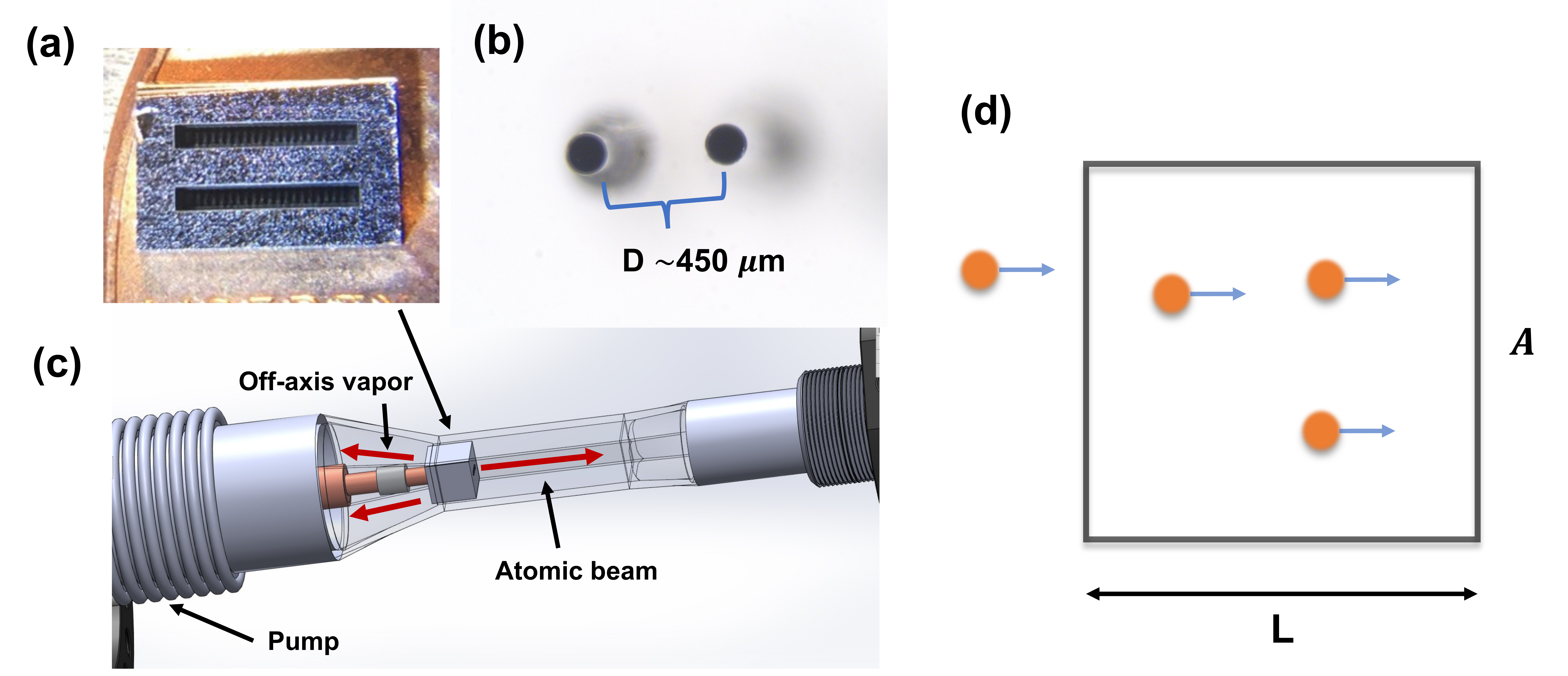}
\caption{(a) The image of cascaded collimator, see Ref \cite{li2019cascaded} for more details. (b) A microscope image of two fiber tips in the two-fiber detector. (c) A 3D model of the vacuum chamber. (d) Diagram of the collection region of a single fiber showing individual atom transits.}
\label{fig:1} 
\end{figure*}

As shown in Fig.\ref{fig:1}(c),
the science chamber is a 12 mm $\times$ 12 mm $\times$ 42 mm cuboid glass cell. Both sides of it are bonded to stainless steel bellows to reduce stress. The right port is connected to a Pfeiffer HICUBE 80 ECO pumping station to maintain a pressure of $10^{-7}$ Torr. The left port is connected to a three-way cross, in which one way is used to insert the atomic oven. The other way is also connected to the same pumping station and assists with pumping the off-axis vapor away.
The copper oven delivers Rubidium vapor into our silicon cascaded collimator (Fig.\ref{fig:1}(a)). 
The principle of this collimator is described in Ref \cite{li2019cascaded}.  The off-axis vapor leaves through the gaps in the collimator while the on-axis atomic beam travels toward the right port.
A small box surrounds the collimator to keep the off-axis vapor away from the interaction region. 

Two multimode (NA=0.22, 105 $\mu$m) fibers are stripped and cleaved to have a clean flat end. They are then attached to a fiber holder which sets their distance to around 450 $\mu$m (Fig \ref{fig:1}(b)). The two-fiber holder is placed at the image plane of our imaging system and aligned to the direction of our atomic beam. The output of each fiber is fed into a single photon detection module (SPCM-AQRH-15). The generated TTL pulses are sent into a Time Interval Analyzer (Guidetech GT668) to be time tagged and stored in the hard disk. The time tags are later used to calculate coincidences.

\section{Single Fiber Second-order correlation theory}

The second-order correlation function $g^{(2)}(\tau)$ is defined to be:
\begin{equation}
g^{(2)}(\tau)= \frac{\langle I_A(t)I_B(t+\tau)\rangle}{\langle I_A(t) \rangle\langle I_B(t+\tau) \rangle} =\frac{\langle n_A(t)/\Delta t\cdot n_B(t+\tau)/\Delta t\rangle}{\langle n_A(t)/\Delta t\rangle \langle n_B(t+\tau)/\Delta t\rangle}
\label{g2}
\end{equation}
Where  $n_{A,B}(t)$ is the number of detected photons from detector A(B) in time bin $\Delta t$ at time t and the intensity $I(t)\propto\frac{n(t)}{\Delta t}$. The effect of $g^{(1)}(\tau)$ can be negligible in our system.
For our thermal atomic beam experiment, the background counts are negligible ($<1\%$). Thus, we ignore the background counts and only consider photons from the atoms.

For an effusive thermal atomic beam with a given output flux, we define a normalized  velocity distribution of the flux $\rho(v)$:
\begin{equation}
    \rho(v)=2\frac{v^3}{v_0^4}e^{-v^2/v_0^2}
\end{equation}
Where $v_0=\sqrt{\frac{2 k_B T}{m}}$, $k_B$ is the Boltzmann constant,  $m$ is the mass of the atom, and $\int_v\rho(v)dv=1$.  This distribution is related to the mean number of atoms that transit our collection region per second $F_N$ (see Fig. \ref{fig:1}(d)) through the formula $n(v)\cdot vA=F_N\rho(v)$, where $n(v)$ is the density of atoms with velocity between $v$ and $v+dv$ and $A$ is the cross-sectional area of the collection region. The relationship between $F_N$ and the average atom number in the field of view $\langle N\rangle$ is then straightforward to calculate:
\begin{equation}
    \langle N\rangle=\int_vA\cdot L\cdot n(v)dv=F_N\int_v \rho(v) \frac{L}{v}dv
    \label{aveN}
\end{equation}
where $L$ is the field of view length along the atomic beam direction.

We calculate the right hand side of Eqn. (\ref{g2}) by separately evaluating the numerator and denominator, starting with the latter.  Here we need the average number of photons detected, which is the product of mean atom number $\langle N\rangle$, scattering rate $R_s$, and collection efficiency $C_{eff}$:
   \begin{equation}
    \langle n_A(t)/\Delta t\rangle=C_{eff} R_s F_N\int_v \rho(v)\frac{L}{v}dv.
    \label{n}
\end{equation}
Moreover, since this average is time-independent, the denominator becomes

\begin{equation}
 \langle n_A(t)/\Delta t\rangle \langle n_B(t)/\Delta t\rangle = \left ( C_{eff}R_sF_N  \int_v \rho(v)\frac{L}{v}dv \right )^2
    \label{denominator}
\end{equation}

To calculate the numerator correctly, we must consider the fluctuating number of atoms in the volume.  If $p(N_f)$ is the probability to have $N_f$ atoms in the field of view, then $\langle N\rangle=\sum_{N_f} p(N_f)\cdot N_f$.  Thus, we can write the numerator as:
\begin{equation}
    \sum_{N_f} p(N_f)\langle(n_{1A}(t)+n_{2A}(t)+...n_{N_fA}(t))/\Delta t\cdot(n_{1B}(t+\tau)+n_{2B}(t+\tau)+...n_{N_fB}(t+\tau))/\Delta t\rangle
    \label{numerator1}
\end{equation}
Where $n_{iA}(t)$ represents the number of emitted by $i^{th}$ atom in time bin $\Delta t$ at time t that reach detector $A$ and similarly for $B$. $n_{iA}(t)\cdot n_{jB}(t+\tau)$ are uncorrelated unless $i=j$. Therefore, we can write (\ref{numerator1}) as the sum of a correlated term and an uncorrelated term:
\begin{equation}
    \sum_{N_f} p(N_f)\cdot \sum_{i=1}^{i=N_f}\langle n_{iA}(t)/\Delta t\cdot n_{iB}(t+\tau)/\Delta t\rangle+\sum_{N_f} p(N_f)\cdot \sum_{i\neq j}\langle n_{iA}(t)/\Delta t\cdot n_{jB}(t+\tau)/\Delta t\rangle
\end{equation}
Since all atoms are equivalent, we may write  $\langle n_i(t)\rangle=\langle n_j(t+\tau)\rangle$.  Therefore, we may calculate everything in terms of just atom 1's emission:
\begin{equation}
    \sum_{N_f} p(N_f) N_f\cdot\langle n_{1A}(t)/\Delta t\cdot n_{1B}(t+\tau)/\Delta t\rangle+\sum_{N_f} p(N_f) N_f(N_f-1) \langle n_{1A}(t)/\Delta t\rangle\cdot \langle n_{1B}(t+\tau)/\Delta t\rangle
    \label{numerator2}
\end{equation}
The relationship between $\langle n_{1A}(t)/\Delta t\rangle$ and $\langle n_A(t)/\Delta t\rangle$ can be derived:
\begin{equation}
\langle n_{A}(t)/\Delta t\rangle=\sum_{N_f} p(N_f)\langle (n_{1A}(t)+n_{2A}(t)+...n_{N_fA}(t))/\Delta t\rangle=\sum_{N_f} p(N_f) N_f \langle n_{1A}(t)/\Delta t\rangle=\langle N\rangle \cdot \langle n_{1A}(t)\rangle
\label{n_ni}
\end{equation}
Using (\ref{denominator}) and(\ref{n_ni}) with (\ref{numerator2}), the numerator can be written as:
\begin{equation}
   \sum_{N_f} p(N_f) N_f\cdot\langle n_{1A}(t)/\Delta t\cdot n_{1B}(t+\tau)/\Delta t\rangle+\sum_{N_f} p(N_f) \frac{N_f(N_f-1)}{\langle N\rangle^2} \left (C_{eff}R_sF_N\int_v \rho(v)\frac{L}{v}dv \right )^2 
    \label{numerator3}
\end{equation}

Next, we are going to solve the first, correlated term, which is generated by the same atom and is related to its second order correlation function.  We will also calculate the required transit time correction for atoms moving through the field of view.
From (\ref{g2}), we know that for a stationary single atom,  $g^{(2)}_{single}(\tau)=\frac{\langle n_{A}'(t)/\Delta t\cdot n_{B}'(t+\tau)/\Delta t\rangle}{\langle n_{A}'(t)\rangle \langle n_{B}'(t+\tau)/\Delta t\rangle}$.  Here $\langle n_{A}'\rangle = \langle n_{BA}'\rangle= C_{eff}R_s \Delta t$ is the mean number of received photons from a stationary atom without transit time correction.  If we then introduce the conditional probability $P(B(\tau)|A)\cdot  \Delta t$ of detecting the second B photon within a time interval $\Delta t$ at time delay $\tau$ given that the first A photon was detected, we obtain
\begin{equation}
    g^{(2)}_{single}(\tau)= \frac{\langle n_{A}'(t)/\Delta t \cdot n_{B}'(t+\tau)/\Delta t\rangle}{C_{eff}R_s\cdot C_{eff}R_s} = \frac{\langle n_{A}'(t)/\Delta t\rangle\cdot P(B(\tau)|A)}{C_{eff}R_s\cdot C_{eff}R_s}=\frac{P(B(\tau)|A)}{C_{eff}R_s}
\end{equation}
For one atom transiting the field of view $L$ with velocity $v$, if a coincidence with a time delay $\tau$ is to be detected, the first photon must have been emitted within a distance $L-v\tau$ to allow the second photon at $\tau$ to be detected. Thus the transit length for the first photon $\langle n_1(t)/\Delta t\rangle$ is effectively reduced to $L-v\tau$, resulting in a correction factor of $\left ( \frac{L-v\tau}{L}\right )$ provided that $\tau<L/v$. No coincidences can be found from the same atom when $v>L/\tau$. 
Combining $g^{(2)}_{single}(\tau)$, the transit time correction factor and Eqn. (\ref{n}), we finally obtain:
\begin{equation}
      \langle n_{1A}(t)/\Delta t\cdot n_{1B}(t+\tau)/\Delta t\rangle =\int_{v=0}^{v=L/\tau} dv C_{eff}R_s \frac{F_N}{\langle N\rangle} \rho(v)\frac{L}{v}\cdot\frac{(L-v\tau)}{L}\cdot C_{eff} R_s\cdot g^{(2)}_{single}(\tau)
    \label{correlated}
\end{equation}
In the above, we may substitute the textbook formula for $g^{(2)}_{single}(\tau)=1-e^{-(3\Gamma/4)\tau}\cdot(\cos(\Omega_\Gamma\tau)+\frac{3\Gamma}{4\Omega_\Gamma}\sin(\Omega_\Gamma\tau))$  \cite{steck2007quantum}, where  $\Omega_\Gamma=\sqrt{\Omega^2-(\frac{\Gamma}{4})^2}$, and $\Omega$ and $\Gamma$ are the Rabi frequency and spontaneous decay rate respectively.

Now we put Eqn. (\ref{correlated}) back into (\ref{numerator3}).  We also assume a Poisson distribution of atom numbers in the field of view, for which 
\begin{equation}
\sum_{N_f}p(N_f)N_f=\langle N\rangle
\end{equation}
and
\begin{equation}
\sum_{N_f}p(N_f)N_f^2=\langle N\rangle^2+\langle N\rangle
\end{equation}
With these formulae, (\ref{numerator3}) becomes:
\begin{equation}
C_{eff}^2R_s^2 F_N\int_{v=0}^{v=L/\tau} \rho(v)\frac{(L-v\tau)}{v}dv\cdot g^{(2)}_{single}(\tau)+\left (C_{eff} R_s  F_N\int_v  \rho(v)\frac{L}{v}dv \right )^2
        \label{numeratorf}
\end{equation}
Combining the denominator (\ref{denominator}) and the numerator (\ref{numeratorf}) we get $g^{(2)}(\tau)$:
\begin{equation}
g^{(2)}(\tau)=\left ( \frac{\int_{v=0}^{v=L/\tau} (1-\frac{v\tau}{L})\frac{\rho(v)}{v}dv}{\int_v  F_N\rho(v)\frac{L}{v}dv\int_v\frac{\rho(v)}{v}dv} \right )\cdot g^{(2)}_{single}(\tau)+1
\end{equation}
From (\ref{aveN}), we can see that the first term in the denominator is actually $\langle N\rangle$. This yields the final expression for $g^{2}(\tau)$:
\begin{equation}
g^{(2)}(\tau)= \left (\frac{\int_{v=0}^{v=L/\tau} (1-\frac{v\tau}{L})\frac{\rho(v)}{v}dv}{\int_v\frac{\rho(v)}{v}dv} \right )\cdot \frac{g^{(2)}_{single}(\tau)}{\langle N\rangle}+1
\end{equation}
The transit time correction $\xi(\tau)$ defined in the main text is the term in parentheses above.  It includes an extra factor of $1/v$ in the integrand compared with the transit time correction derived in Ref \cite{kimble1978multiatom}.  Conceptually, it is because slower atoms contribute more photons per transit and thus have a higher weight in the $g^2(\tau)$. Monte Carlo wave function simulation is also implemented, and our transit time correction factor fits well with the simulation result.


We then fit this formula to our 78 $\degree$C thermal atomic beam data.
The averaged atom number $\langle N \rangle$, the field of view $L$, and the Rabi frequency $\Omega$ in $g^2_{single}(\tau)$ are fitted to the data while $\rho(v)$ is the 78 $\degree$C atomic beam Maxwell-Boltzmann velocity distribution. Because of the intensity variance in the collecting region, the $g^2_{single}(\tau)$ is averaged over a Gaussian distributed Rabi frequency $\Omega$. The fitted parameters are $\langle N \rangle=0.138$, $L=25\ \mu$m and $\Omega$ is a Gaussian distribution with $\mu=6\Gamma$, $\sigma=1.5\Gamma$. The theory curve together with the experimental data is shown in Fig.4 (a) in the main text. 


\section{Two-fiber velocity detector data processing}
Similar to the single fiber second-order correlation theory, here we start with a formula for the coincidences distribution $C(\tau)d\tau$ in the time domain and convert the coincidences into the velocity domain later.
As before, we ignore the background counts and only consider photons from the atomic beam. We set the field of view of the fiber to be $d_f$ and the distance between two fibers in the objective plane to be $d$. We make the approximation that $d_f/d\ll1$.
Two components contribute to the coincidences:
\begin{equation}
    C(\tau)d\tau=\mathrm{uncorrelated\_term} + \mathrm{correlated\_term}
    \label{coin}
\end{equation}
\begin{figure*}[b]
\centering
\includegraphics[width=0.5\linewidth]{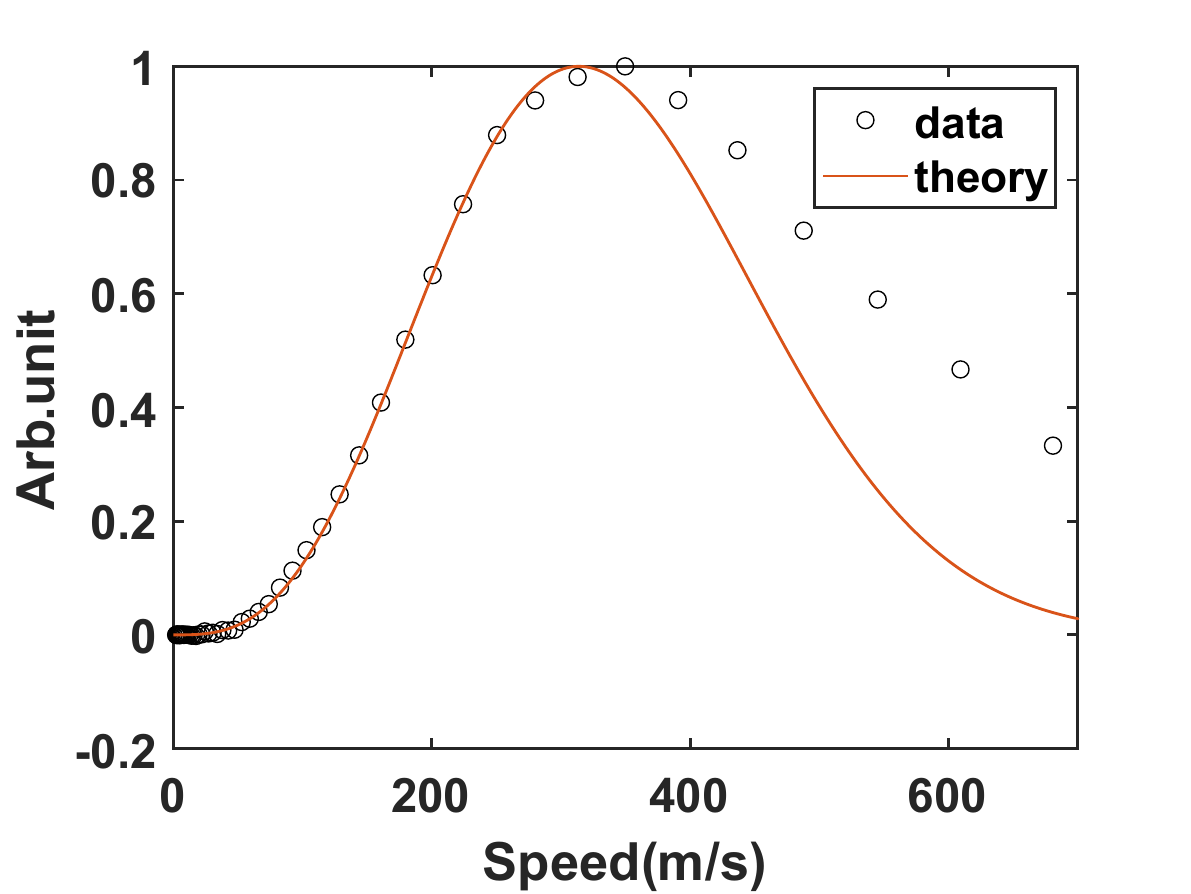}
\caption{70 $\degree$C thermal atomic beam experimental data after processing versus theory. The distance between two fibers in the objective plane $d$=55 $\mu$m.}
\label{fig:calib} 
\end{figure*}
The first uncorrelated term is the accidental coincidences generated by randomly having atoms at fiber A and atoms at fiber B at the same time. This term has no relationship with time delay $\tau$ and can be written as:
\begin{equation}
    \mathrm{uncorrelated\_term}=
    \int_v C_{eff}R_sF_N\rho(v)\frac{d_f}{v}dv\cdot\int_v C_{eff}R_sF_N\rho(v)\frac{d_f}{v}dv d\tau
    \label{uncorrelated}
\end{equation}
Where $d\tau$ is the size of time bins for coincidences
The correlated term comes from atoms with velocity $v_\tau=\frac{d\pm d_f}{\tau}\approx\frac{d}{\tau}$. We ignore $d_f$ here since $\frac{d_f}{d}$ is small, these atoms emitted photons in fiber A and in fiber B at $\tau$ later:
\begin{equation}
    C_{eff}R_s\frac{d_f}{v_\tau}\cdot C_{eff}R_s\frac{d_f}{v_\tau}F_N\rho(v_\tau)dv_\tau
    \label{extra_term}
\end{equation}
Since $v_\tau=\frac{d}{\tau}$, put $dv_\tau=d\tau\cdot\frac{v_\tau^2}{d}$ into (\ref{extra_term}) we get:
\begin{equation}
        \mathrm{correlated\_term}=  C_{eff}^2R_s^2\frac{d_f^2}{d}F_N\rho(v_\tau)d\tau
\end{equation}
If we divided the uncorrelated term (\ref{uncorrelated}) on both side of (\ref{coin}), combine the definition of average atom number$\langle N\rangle$(\ref{aveN}) we get:
\begin{equation}
    C'(\tau)=1+\frac{\frac{d_f}{d}\rho(v_\tau)}{\langle N\rangle\int_v\frac{\rho(v)}{v}dv}
    \label{twofiber}
\end{equation}

This formula shows that it needs to be in the single atom regime ($\langle N\rangle\ll 1$) so that the second correlated term is large enough to be detected.

We can also see that when $\tau\rightarrow\infty$, $C'(\tau)\rightarrow 1$. Since $g_{AB}^{(2)}(\tau)$ is also the coincidences distribution normalized to infinity time delay, $C'(\tau)$ equals the cross-correlation $g_{AB}^{(2)}(\tau)$. Next, as we mentioned in the paper, that value $g_{AB}^{(2)}(\tau)-g_{AB}^{(2)}(\infty)$, after normalization, is the coincidences probability density from atoms in time domain $n_{AB}(\tau)$, which is proportional to $\rho(v_\tau)$. Given $n_{AB}(\tau)d\tau=n_{AB}(v)dv$ and $\tau=\frac{d}{v}$, we can get the coincidences probability density in velocity space $n_{AB}(v)=n_{AB}(\tau)\cdot\frac{d}{v^2}$. And finally we can get $n_{AB}(v)\propto\rho(v)\cdot\frac{1}{v^2}$.\\

To calibrate our theory with data, we measured the unfiltered thermal atomic beam at 70 $\degree$ C with our two-fiber detector. The result is shown in Fig. \ref{fig:calib}. The circles are the experimental data after processing, and the red curve is the theoretical curve for 70 $\degree$C atomic beam Maxwell Boltzmann distribution. The theory fits very well for velocities below 300 m/s which is the range we focused on. The error becomes larger when velocity is larger because $d_f/d\approx0.45$ and it could have an uncertainty error around 22.5\%. Also, because of the imperfect imaging, some atoms can emit photons into both fibers during the transit from fiber A to fiber B, which causes some spurious population at high velocity.

For $\Delta=-20$ MHz and $\Delta=-10$ MHz, as mentioned in the main text, the fast unpumped atoms and vapor are more than the selected atoms. Thus we also measured the $g^{(2)}_b(\tau)$ with the pump beam and probe beam only to get the contributions from fast unpumped atoms and vapor. Then, we subtracted the background contribution when calculating $\rho(v)$ with the selected atoms according to (\ref{twofiber}) and assuming $\int_v\frac{\rho(v)}{v}dv$ have roughly the same value. 
\begin{figure*}[hb]
\centering
\includegraphics[width=0.9\linewidth]{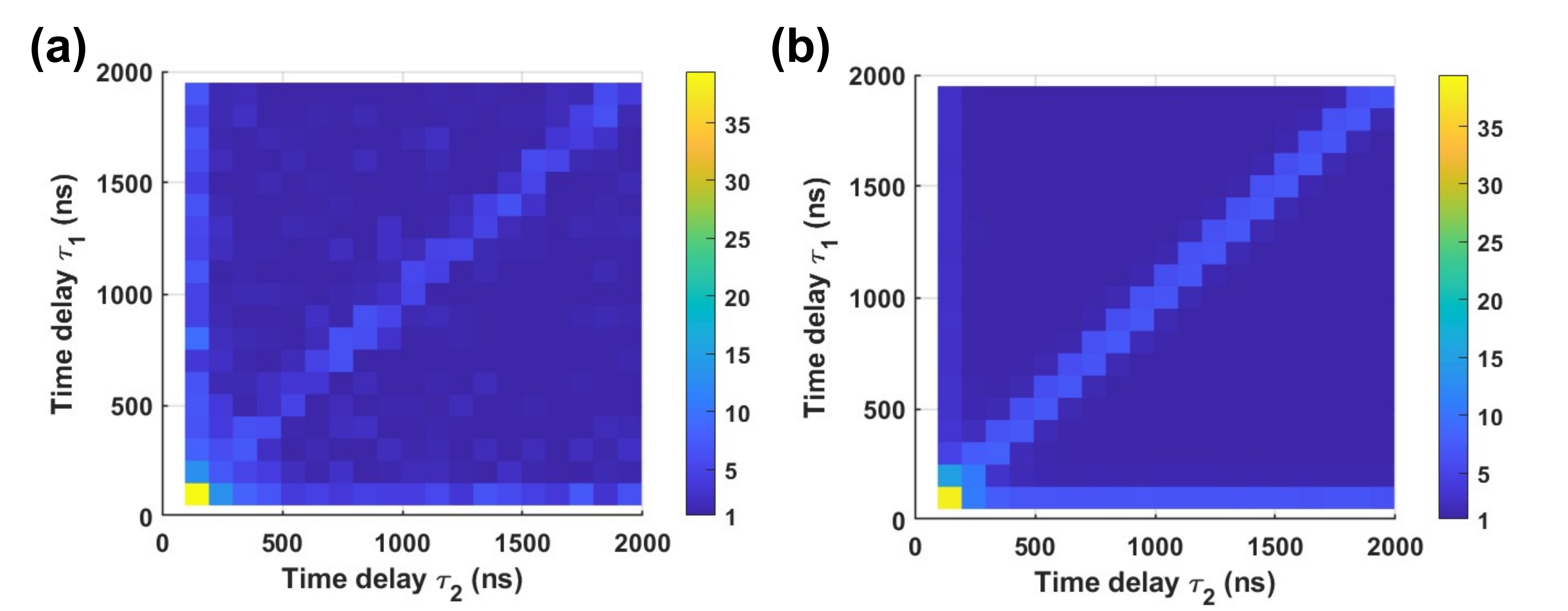}
\caption{(a) The experimental $g^3(\tau_1,\tau_2)$ value for 78 $\degree$C unfiltered thermal atomic beam. Same as the linear scale of Figure 5 (a) in the main text.  (b) The simulated $g^3(\tau_1,\tau_2)$. The time bin size is 100 ns $\times$ 100 ns}
\label{fig:g3} 
\end{figure*}

\section{Monte Carlo wave function simulation}
Monte Carlo wave function (MCWF) simulation is also implemented to test our theory. 
The Monte Carlo wave function (MCWF) simulation  was designed to mimic what happened in our experimental system to calculate the $g^{(3)}(\tau_1,\tau_2)$. The atoms are generated according to the Poisson distribution, and the velocities are chosen from the 78 $\degree$C atomic beam Maxwell Boltzmann distribution. The atoms then fly into a laser beam and interact with it. The wavefunctions are evolved according to the MCWF procedure \cite{molmer1993monte}.
When atoms are within the field of view of the fiber, their emitted photons' time tags are registered and stored. Then the same algorithm used to calculate experimental $g^{(3)}(\tau_1,\tau_2)$ is used on the simulation data. 

\begin{figure*}[h]
\centering
\includegraphics[width=\linewidth]{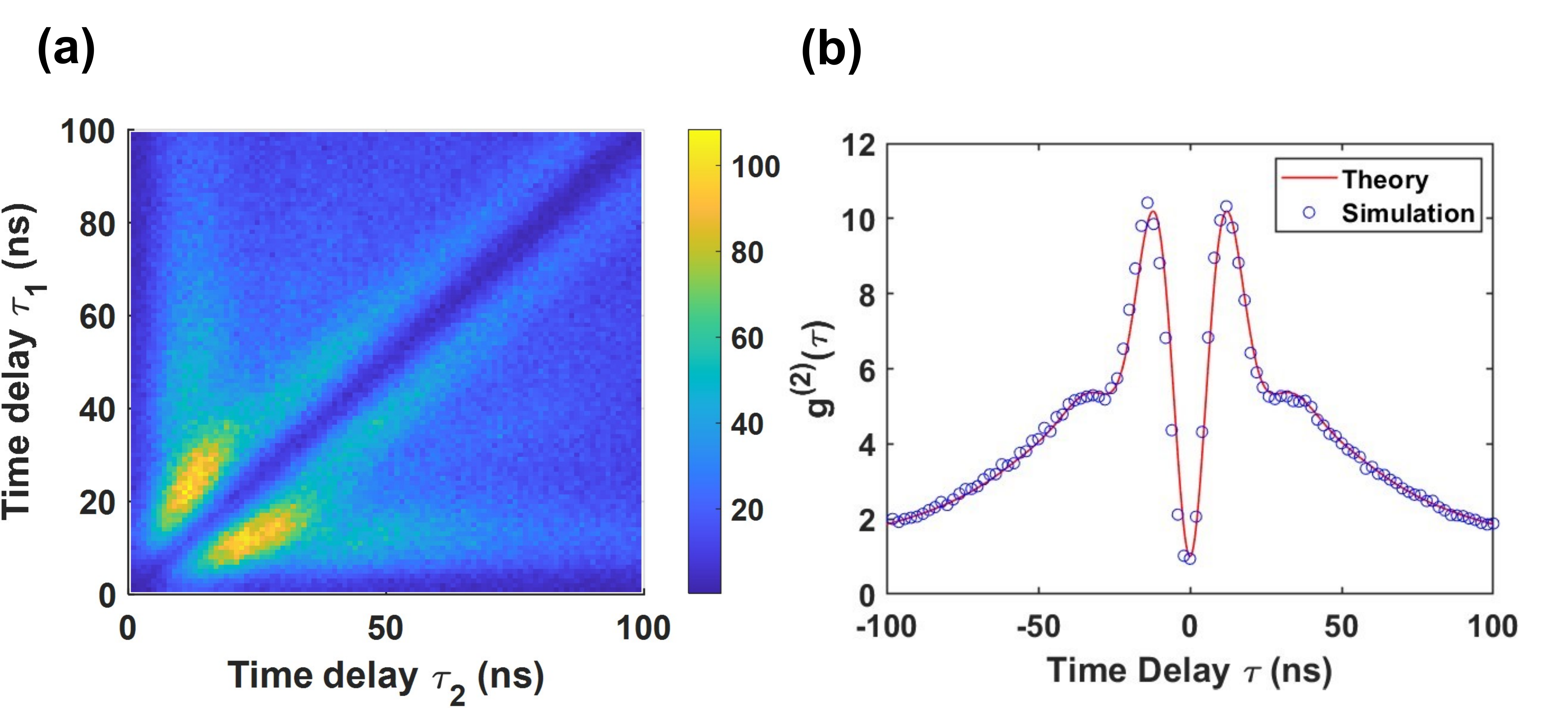}
\caption{(a) The simulated $g^{(3)}(\tau)$ with 1 ns $\times$ 1 ns time bins to see the dynamics near zero time delay. (b) The simulated $g^{(2)}(\tau)$ with the same parameter we used to fit our experimental data, plotted together with our theoretical curve. $\langle N \rangle=0.138$, $L=25\ \mu$m and $\Omega$ is a Gaussian distribution with $\mu=6\Gamma$, $\sigma=1.5\Gamma$}
\label{fig:g2_simulated} 
\end{figure*}

The result is shown in Fig. \ref{fig:g3}. The colorbar is in linear scale and the simulation fits the data quite well. Because of the limited three-photon coincidences rate,  figure  \ref{fig:g3} and the $g^{(3)}(\tau_1,\tau_2)$ in our main text all used 100 ns $\times$ 100 ns time bins to have more averaging effect to overcome the shot noise. The resolution is limited, and the dynamics inside the 100 ns $\times$ 100 ns are averaged out. Thus, MCWF simulations were done to calculate the $g^{(3)}(\tau_1,\tau_2)$ with no dead time and with 1 ns $\times$ 1 ns time resolution near zero time delay (Fig. \ref{fig:g2_simulated} (a)). We can see that similar to $g^{(2)}(0)$,$g^{(3)}(0,0)=1$ and when $\tau_1=0$, $\tau_2=0$ or $\tau_1=\tau_2$, $g^{(3)}(\tau_1,\tau_2)$ equals 1. Then, $g^{(3)}(\tau_1,\tau_2)$ has a huge peak where three consecutive photons from the same atom create coincidences. With 100 ns resolution, this dynamics is averaged into one bunching peak around zero time delay.
In the future, we can put some graphite in the system to absorb the accumulated vapor and average for a much longer time to reduce the three-photon-coincidences shot noise. Then, we can use three detectors to eliminate the deadtime constraint and use a smaller time bin (4 ns) to see the dynamics near zero time delay.

Similar simulations are also done for the $g^{(2)}(\tau)$, and it agrees well with our theory in section II (see Fig. \ref{fig:g2_simulated}).

\bibliography{references}